\documentclass[iop]{emulateapj}
\usepackage{apjfonts}
\usepackage{natbib}
\usepackage{graphicx}
\usepackage{times}
\usepackage{xspace} 
\usepackage{booktabs}

\slugcomment{\it Submitted --- 2020 April 5; Published --- 2020 June 1}
\shorttitle{Einstein Ring}
\shortauthors{Stern \& Walton}

\def\mg{{MG~1131+0456}}

\def\ie{{i.e.}}
\def\eg{{e.g.}}

\def\wise{{\it WISE}}

\def\deg{\ifmmode {^{\circ}}\else {$^\circ$}\fi}
\def\kms{\ifmmode {\rm\,km\,s^{-1}}\else
    ${\rm\,km\,s^{-1}}$\fi}
\def\ergcm2s{\ifmmode {\rm\,erg\,cm^{-2}\,s^{-1}}\else
    ${\rm\,erg\,cm^{-2}\,s^{-1}}$\fi}
\def\ergAcm2s{\ifmmode {\rm\,erg\,cm^{-2}\,s^{-1}\,\AA^{-1}}\else
    ${\rm\,erg\,cm^{-2}\,s^{-1}\,\AA^{-1}}$\fi}
\def\ergs{\ifmmode {\rm\,erg\,s^{-1}}\else
    ${\rm\,erg\,s^{-1}}$\fi}

\def\spose#1{\hbox to 0pt{#1\hss}}
\def\simlt{\mathrel{\spose{\lower 3pt\hbox{$\mathchar"218$}}
     \raise 2.0pt\hbox{$\mathchar"13C$}}}
\def\simgt{\mathrel{\spose{\lower 3pt\hbox{$\mathchar"218$}}
     \raise 2.0pt\hbox{$\mathchar"13E$}}}

\def\plotfiddle#1#2#3#4#5#6#7{\centering \leavevmode \vbox to#2{\rule{0pt}{#2}} \includegraphics{#1}}

\begin{document}

\title{A Redshift for the First Einstein Ring, MG~1131+0456}

\author{Daniel~Stern\altaffilmark{1} \&
Dominic~J.~Walton\altaffilmark{2}}

\altaffiltext{1}{Jet Propulsion Laboratory, California Institute of Technology, 4800 Oak Grove Drive, Pasadena, CA 91109, USA [e-mail: daniel.k.stern@jpl.nasa.gov]}

\altaffiltext{2}{Institute of Astronomy, University of Cambridge, Madingley Road, Cambridge CB3 0HA, UK}

\begin{abstract} 

\mg\ is a radio-selected gravitational lens, and is the first known Einstein ring.  Discovered in 1988, the system consists of a bright ($S_{\rm 74~MHz} = 3.7~{\rm Jy}$) radio source imaged into a ring and two compact, flat-spectrum components separated by 2\farcs1. The ring is optically faint ($R = 23.3$), rising steeply into the near- and mid-infrared ($K = 17.8$; $W2 = 13.4$). The system has been intensively studied in the intervening years, including high-resolution radio imaging, radio monitoring, and near-infrared imaging with {\it Hubble} and Keck.  The lensing galaxy is at $z_{l} = 0.844$.  However, to date, no spectroscopic redshift had been reported for the lensed source.  Using archival Keck data from 1997, we report the robust detection of a single narrow emission line at 5438~\AA\ which we associate with \ion{C}{3}]~$\lambda 1909$~\AA\ from a type-2 quasar at $z_{s} = 1.849$.  Support for this redshift identification comes from weaker emission associated with \ion{C}{4}~$\lambda 1549$~\AA\ and \ion{He}{2}~$\lambda 1640$~\AA, typical of type-2 quasars, as well as the lack of emission lines in archival near-infrared Keck spectroscopy.  We also present, for the first time, Cycle~1 {\it Chandra} observations of \mg, which clearly resolves into two point sources with a combined flux of $\sim 10^{-13}\, {\rm erg}\, {\rm cm}^{-2}\, {\rm s}^{-1}$ and a best-fit column density of $\sim 3 \times 10^{22}\, {\rm cm}^{-2}$.  We suggest a new method to identify candidate lensed AGN from low-resolution X-ray surveys such as eROSITA by targeting sources that have anomalously high X-ray luminosity given their mid-infrared luminosity.

\end{abstract}


\section{Introduction}

\mg\ was the first discovered Einstein ring \citep{Hewitt:88}, identified shortly after the first discovery of lensed arcs in rich galaxy clusters \citep{Lynds:87, Soucail:87} and about a decade after the first discovery of a lensed quasar \citep{Walsh:79}.  The unusual radio structure consists of an elliptical ring with major and minor axes of 2\farcs2 and 1\farcs6, respectively, accompanied by a pair of compact sources slightly offset from the ring ($\sim 0\farcs3$ to the southwest).  A low surface brightness component is also evident to the southwest of the ring.  Shortly after its discovery, \mg\ was successfully modelled by \citet{Kochanek:89} to be a gravitationally lensed radio galaxy, where the radio jet aligns with the astroid caustic of the foreground lens and produces an Einstein ring \citep{Einstein:36}, the radio core is slightly misaligned with the lensing galaxy, producing the pair of compact sources, and the low surface brightness component to the southwest is a radio lobe. \citet{Chen:93} and \citet{Hewitt:95} present improved, multifrequency radio images of \mg, showing the compact sources to be flat spectrum sources, as expected for radio cores, and that these compact sources are weakly variable. The radio flux rises to lower frequencies, with $S_{\rm 74~MHz} = 3.7 \pm 0.4$\, Jy \citep{Cohen:07}.

The system (i.e., lens plus source) is faint at optical wavelengths, $V = 23.0$, and rises steeply in the near-infrared, $K = 16.0$ \citep{Annis:92}.  Separating the lens from the source reveals the background Einstein ring to be extremely red, with $R = 23.3$ and $K = 17.8$ \citep{Annis:92}. Indeed, \citet{Larkin:94} noted \mg\ was among the first of a new class of extremely red objects \citep[$R - K \simgt 6$; e.g.,][]{Elston:88}, and suggested the red colors might be due to dust extinction in the lensing galaxy. However, they also highlighted various objections to that hypothesis; e.g., the lensing galaxy appeared to be early-type, and such galaxies typically do not contain much dust, particularly on the $\geq 10$~kpc scale separating the lensing paths of the background source.  Based on near-infrared imaging from the {\it Hubble Space Telescope} and revised multi-wavelength lens modeling of this system, \citet{Kochanek:00} demonstrated that the host galaxy of \mg\ is an intrinsically red galaxy, as suggested by \citet{Annis:93}, and that tidal perturbations from nearby galaxies can explain the elongated shape of the Einstein ring, as suggested by \citet{Chen:95}.

Based on optical to near-infrared imaging and modeling of the gravitational lens, several authors concluded that the lens was an early-type galaxy at $z_{l} = 0.7 - 0.9$ \citep[e.g.,][]{Larkin:94, Kochanek:00b}, consistent with the tentative $z_l = 0.85$ found by \citet{Hammer:91}.  \citet{Tonry:00} present an optical spectrum of the system obtained with the Keck~2 telescope, conclusively demonstrating that the lensing galaxy is an early-type galaxy at $z_{l} = 0.844$ with [\ion{O}{2}]~$\lambda \lambda 3727$~\AA\ emission, CaHK absorption, and a strong D4000 break.  From the lens model, they argue that the source redshift for \mg\ is at $z_{s} > 1.9$.  However, prior to this Letter, no spectroscopic redshift for the Einstein ring had been reported.

Here we report on a re-analysis of archival Keck observations of \mg\ which provide a spectroscopic redshift for the lensed radio galaxy.  We also present archival observations with {\it Chandra} (\citealt{CHANDRA}) that resolve the source into two bright X-ray point sources, associated with the lensed radio core.  Throughout this paper, magnitudes are reported in the Vega system.  When computing luminosities, we adopt the $\Lambda$CDM concordance cosmology: $H_0 = 70\, {\rm km}\ {\rm s}^{-1}\, {\rm Mpc}^{-1}$, $\Omega_{\rm M} = 0.3$, and $\Omega_\Lambda = 0.7$.

\section{Observations \& Analysis}

We re-discovered \mg\ while searching for lensed obscured quasars. While over two hundred lensed quasars are currently known\footnote{As of 2020 March 14, the Gravitationally Lensed Quasar Database, https://web1.ast.cam.ac.uk/ioa/research/lensedquasars, contains 219 known lensed quasars.}, the number of known lensed obscured, or type-2, quasars is significantly smaller, potentially in the single digits despite the fact that most accreting black holes in the universe are obscured.  Examples include MG~2016+112 at $z_{s} = 3.273$ \citep{Lawrence:84}, MG~J0414+0534 at $z_{s} = 2.639$ \citep{Hewitt:92}, and PMN~J0134$-$0931 at $z_{s} = 2.216$ \citep{Winn:02}. Notably, all of these were radio selected.  Mimicking the long delay between the discovery of distant, radio-loud, type-2 quasars \citep[i.e., high-redshift radio galaxies; e.g.,][]{Spinrad:82} and the discovery of distant, radio-quiet, type-2 quasars \citep[e.g.,][]{Stern:02a}, only more recently have lensed, radio-quiet, type-2 quasars been identified, with examples including MACS~J212919.9$-$074218 at $z_{s} = 1.212$ \citep{Stern:10b} and W2M~J1042+1641 at $z_{s} = 2.517$ \citep{Glikman:20}.  Both those discoveries were serendipitous.

In order to find additional lensed obscured quasars, we correlated known galaxy-galaxy lenses from the Master Lens Database (Moustakas et al., in preparation) with mid-infrared photometry from the {\it Wide-field Infrared Survey Explorer} \citep[\wise;][]{Wright:10}. We then identified which systems included an active galaxy based on red colors between the first two \wise\ bands, $W1$ (3.4~$\mu$m) and $W2$ (4.6~$\mu$m); namely, $W1 - W2 \geq 0.8$ is a highly effective and efficient criterion for identifying both obscured and unobscured quasars \citep[e.g.,][]{Stern:12, Assef:13}.  Given the $\sim 6\arcsec$ resolution of \wise, this selection could identify either lensed quasars or systems where a quasar is responsible for the lensing; both scenarios are scientifically valuable \citep[e.g.,][]{Courbin:12}.  Only one high-quality galaxy-galaxy lens (i.e., ``lensgrade'' = A) had such red \wise\ colors:  \mg.

The re-discovery of this interesting, bright, mid-infrared source ($W2 = 13.40$) with extremely red \wise\ colors ($W1 - W2 = 0.93$) inspired a search of the archives to learn more about this source that had attracted significant attention after its discovery three decades ago, but had largely been forgotten for the past two decades.

% TABLE 1:  KECK OBSERVATIONS
% \scriptsize
\begin{deluxetable}{llcccl}
% \tablewidth{0pt}
\tablecaption{Archival Keck spectroscopy of \mg.}
\tablehead{
\colhead{} &
\colhead{} &
\colhead{PA} &
\colhead{Exposure} &
\colhead{$\lambda \lambda$} &
\colhead{} \\
\colhead{UT Date} &
\colhead{Inst.} &
\colhead{(deg)} &
\colhead{(s)} &
\colhead{($\mu$m)} &
\colhead{PI}}
\startdata
1997 Mar 31    & LRIS    &  45 & 3300 & 0.37 - 0.87 & Tonry \\
1999 May 11    & LRIS    &  63 & 5400 & 0.68 - 0.84 & Tonry \\
2000 Apr 12    & NIRSPEC & 116 & 1800 & 1.14 - 1.38 & Larkin \\
2000 Apr 12    & NIRSPEC & 117 &  600 & 2.00 - 2.38 & Larkin \\
2002 Feb 08    & ESI     &  93 & 5400 & 0.39 - 1.02 & Koopmans \\
2007 Jan 27    & NIRSPEC &  60 & 2400 & 1.43 - 1.81 & Canalizo \\
2007 Jan 27    & NIRSPEC &  60 & 2400 & 1.95 - 2.30 & Canalizo 
% 2020 Mar 10  & NIRES   &     &      &             & Urry
\enddata
\label{table:observations}
\tablecomments{$\lambda\lambda$ lists the wavelength range covered
by the observations.}

\end{deluxetable}
% \normalsize

%
% FIGURE 1 - OPTICAL SPECTRUM
\begin{figure}
\plotfiddle{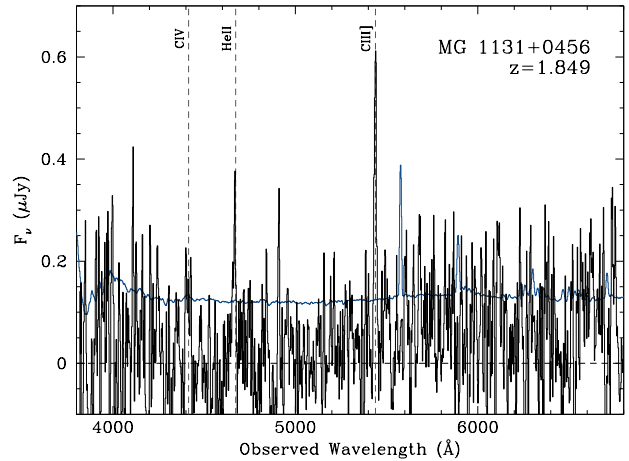}{2.6in}{-90}{32}{32}{-120}{190}
\caption{Keck/LRIS observation of \mg\ from March 1997 {\bf in black; the blue line shows the 1$\sigma$ error spectrum.  Spectra were extracted with a 1\farcs0 width aperture and smoothed with a 3-pixel boxcar.}  A strong emission line is present at 5438~\AA, associated with \ion{C}{3}]~$\lambda 1909$~\AA, with supporting weaker lines associated with \ion{C}{4}~$\lambda 1549$~\AA\ and \ion{He}{2}~$\lambda 1640$~\AA\ also detected.  The spectrum is typical of type-2 quasars.}
\label{fig:spectra}
\end{figure}

\subsection{Keck}

% Like many other high-redshift radio galaxies from the 5~GHz MIT-Green Bank (MG) survey \citep{Bennett:86, Lawrence:86}, 

\mg\ was observed by the Keck telescopes in the late 1990s.  Examples of similar Keck observations of high-redshift radio galaxies in that era appear in \citet{Stern:99f}.  Table~\ref{table:observations} presents a list of all publicly available Keck spectroscopy of \mg\ in the Keck archive.  Other than the 1999 observations which appear in \citet{Tonry:00}, these data have not previously appeared in the literature.

We have analyzed all these data.  Continuum is detected in all cases, coming from the lens for the optical data and primarily from the background radio galaxy source for the near-infrared data.  As noted earlier, the Low Resolution Imaging Spectrograph \citep[LRIS;][]{Oke:95} clearly reveals the lens to be an early-type galaxy with [\ion{O}{2}] emission \citep{Tonry:00}.  

The only observation in which a spectral feature is detected from the source is the first observation, obtained with LRIS in 1997. Those data, {\bf presented in Figure~\ref{fig:spectra}}, were obtained through the 1\farcs0 slit at a position angle $PA = 45^\circ$ with the 300~$\ell\ {\rm mm}^{-1}$ grating ($\lambda_{\rm blaze} = 5000$~\AA; resolving power $R \equiv \lambda / \Delta \lambda \approx 600$ for objects filling the slit).  This $PA$ is slightly offset from the $63^\circ$ angle of the radio cores.  At the time, LRIS was a single-beam spectrograph; a blue arm has since been added. Two observations were obtained, with exposure times of 1500~s and 1800~s. We processed the spectra using standard techniques, and flux calibrated the combined, extracted spectrum using an archival sensitivity function obtained through the same grating.

An emission line at 5438~\AA\ is well detected in both individual exposures, offset by 1\arcsec\ from the lensing galaxy as expected given the geometry of the Einstein ring and radio cores described in \S~1.  No continuum is detected from the lensed radio galaxy, and a weaker emission line is detected at 4667~\AA.  As shown in Figure~\ref{fig:spectra}, we associate these lines with \ion{C}{3}]~$\lambda 1909$~\AA\ and \ion{He}{2}~$\lambda 1640$~\AA\ at $z_{s} = 1.849 \pm 0.002$ (the uncertainty is a conservative estimate, reflecting statistical uncertainties, systematic wavelength calibration uncertainties, and kinematics in the source).  Another weak feature is also evident in the extracted spectrum corresponding to \ion{C}{4}~$\lambda 1549$~\AA. These high-ionization, narrow, high equivalent width emission lines are typical of luminous, obscured active galactic nuclei (AGN), such as high-redshift radio galaxies \citep[\eg,][]{Stern:99a} and type-2 quasars \citep[\eg,][]{Stern:02a}.  As detailed below, the redshift we derive for \mg\ places other key emission features at wavelengths challenging or inaccessible from the ground.

At observed optical wavelengths, we might have expected to see Ly$\alpha$ and the \ion{Mg}{2}~$\lambda \lambda 2800$~\AA\ doublet. Ly$\alpha$, typically one of the brightest features in an active galaxy, is expected at 3464~\AA. Though accessible from the ground, none of the Keck observations to date reach such a blue wavelength (see Table~\ref{table:observations}). \ion{Mg}{2} is expected at 7977 \AA, which is a wavelength that was sampled by multiple archival observations, albeit in a noisier part of the spectrum due to telluric OH emission.  In the composite MG radio galaxy spectrum of \citet{Stern:99a}, \ion{C}{3}] and \ion{Mg}{2} have comparable strength, though the composite radio galaxy spectrum presented in \citet{McCarthy:93} has \ion{C}{3}] roughly twice as strong as \ion{Mg}{2}, and the Seyfert~II composite from \citet{Ferland:86} has \ion{C}{3}] three times as strong as \ion{Mg}{2}.  We ascribe the non-detection of \ion{Mg}{2} from the lensed galaxy to a combination of a more Seyfert~II-like line ratio and elevated noise due to telluric emission at redder optical wavelengths.  {\bf Finally, we note that the 2002 observations with the Echellette Spectrograph and Imager \citep[ESI;][]{Sheinis:02} do not detect the \ion{C}{3}] line, likely due to a combination of a sub-optimal position angle for those observations and the line being at a wavelength where the light is split across two orders of this cross-dispersed spectrograph.}

At observed near-infrared wavelengths, we might have expected to
see [\ion{O}{2}], H$\beta$~$\lambda 4861$~\AA, [\ion{O}{3}]~$\lambda \lambda 4959, 5007$~\AA, and/or H$\alpha$~$\lambda 6563$~\AA. [\ion{O}{2}] is expected at $1.06~\mu$m, which is bluewards of the NIRSPEC observations.  The H$\beta$/[\ion{O}{3}] complex is redshifted to 1.38 - 1.42~$\mu$m, which is between the $J$- and $H$-bands, and in a wavelength range strongly affected by atmospheric absorption. H$\alpha$, usually one of the strongest lines in a galaxy or active galaxy, is redshifted to $1.87~\mu$m, which is between the $H$- and $K$-bands, and is also in a wavelength range strongly affected by atmospheric absorption.  The non-detection of spectral features in archival near-infrared Keck spectra which clearly detect strong continuum lends further support to our redshift assignment and likely explains why a redshift had not previously been reported for \mg\ despite multiple observations over decades.

\subsection{Chandra}

\begin{figure}
\plotfiddle{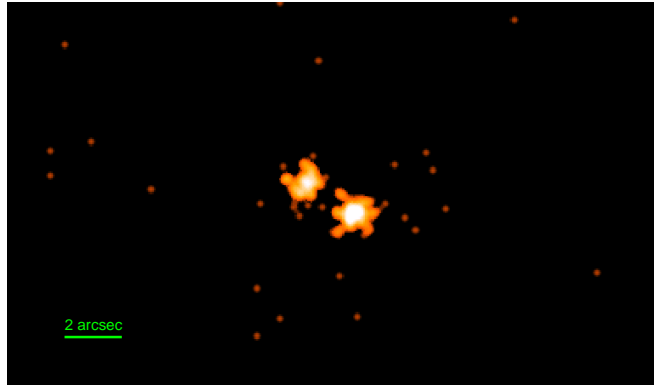}{2.2in}{0}{32}{32}{-120}{8}
\caption{The \textit{Chandra} X-ray image of \mg, combining OBSIDs 423 and 424.  North is up and east is to the left. \mg\ is clearly resolved into two point sources, image A to the north-east and image B to the south-west.}
\label{fig:chandra}
\end{figure}

\textit{Chandra} observed \mg\ with its ACIS-S detector (\citealt{Garmire:03}) on two occasions beginning less than a year after launch (OBSIDs 423 and 424, taken on 2000 May 2 and 2000 December 15, respectively), each with an exposure of $\sim$7\,ks. We reprocessed both observations with {\small CIAO} v4.11 and the latest \textit{Chandra} calibration files. Cleaned event files were generated with the {{\small CHANDRA\_REPRO}} script, with the {{\small EDSER}} sub-pixel event re-positioning algorithm enabled (\citealt{CHANDRA_EDSER}). Following previous work (\eg, \citealt{Reynolds14}), the cleaned event files were re-binned to 1/8th of the ACIS pixel size before smoothing with a Gaussian (0\farcs25 FWHM) for visualisation and to define our source regions. \mg\ is clearly resolved into two X-ray sources separated by 2\arcsec\ by \textit{Chandra} (Figure \ref{fig:chandra}), similar to the two radio cores reported by \citet{Hewitt:88}, one to the north-east (image A) and one to the south-west (image B). For each epoch, we initially extracted individual spectra from each of the two images with the {\small SPECEXTRACT} script, which also generated the relevant instrumental response files. Source events were extracted from circular regions of radius 1\farcs25 (ensuring the two extraction regions do not overlap), while background was estimated from much larger regions of blank sky on the same chip as \mg\ (radius 27\arcsec).

% \citealt{Reis14nat, Reynolds14, Walton15lqso}

% FIGURE 2 - X-RAY IMAGE

We modeled these four spectra with a simple absorbed powerlaw continuum, allowing for absorption both from our own Galaxy (with the column fixed to $N_{\rm{H, Gal}} = 2.87 \times 10^{20}$\,cm$^{-2}$; \citealt{NH}) and intrinsic to the source (\ie, at $z_s = 1.849$), adopting the solar abundance set reported in \cite{tbabs}. Each spectrum was grouped to a minimum of 1 count per bin, and the data were analyzed by reducing the Cash statistic (\citealt{cstat}); the \textit{Chandra} data are modeled over the 0.35--8.0\,keV bandpass in the observed frame, corresponding to rest-frame energies of $\sim$1--23\,keV. The spectral parameters are consistent for all four of the individual spectra, although we note that the signal-to-noise (S/N) of each is fairly poor, as they are comprised of only 26--56 counts ($165 \pm 13$ counts in total, combining both images and both epochs). We also find that, within the limitations of the available data and assuming common spectral parameters, both the image flux ratios and the combined image fluxes are consistent between the two epochs. In the X-ray band, image B is $1.5 \pm 0.4$ times brighter than image A, consistent with the radio core flux ratios of $1.23 \pm 0.10$ at 15~GHz and $1.38 \pm 0.22$ at 22~GHz found by \citet{Chen:93}.

Given this, we combined the data from both images and both epochs to produce an average spectrum for \mg\ (using {\small ADDASCASPEC}). Fitting this data with the above absorbed powerlaw model, we find an intrinsic absorption column of $N_{\rm{H}} = 3.0^{+1.7}_{-1.5} \times 10^{22}$\,cm$^{-2}$ and a photon index of $\Gamma = 1.7^{+0.4}_{-0.3}$. The latter is a  typical X-ray continuum slope for AGN.  For example, based on a study of the broadband X-ray properties of 838 AGN from the 70-month {\it Swift}/BAT all-sky survey, \citet{Ricci:17} find median values of $\Gamma = 1.78 \pm 0.01$ for non-blazar AGN and $\Gamma = 1.54 \pm 0.05$ for flat-spectrum radio quasars. We also compute the combined 2--10\,keV flux (rest-frame) of the two images, mimicking what would be seen by observatories that do not have the spatial resolution of \textit{Chandra}. After correcting for the line-of-sight absorption,  we find $F_{\rm{A+B; 2-10}} = (5.5 \pm 1.0) \times 10^{-14}$\,\ergcm2s. Similarly, the total observed flux (prior to any absorption correction) of the combined images over the full \textit{Chandra} band (0.35--8\,keV observed frame, 1--23\,keV rest-frame) is $F_{\rm{A+B; 1-23}} = (8.7 \pm 1.8) \times 10^{-14}$\,\ergcm2s.

% FIGURE 3 - L(X) - L(6um) RELATION
\begin{figure}
\plotfiddle{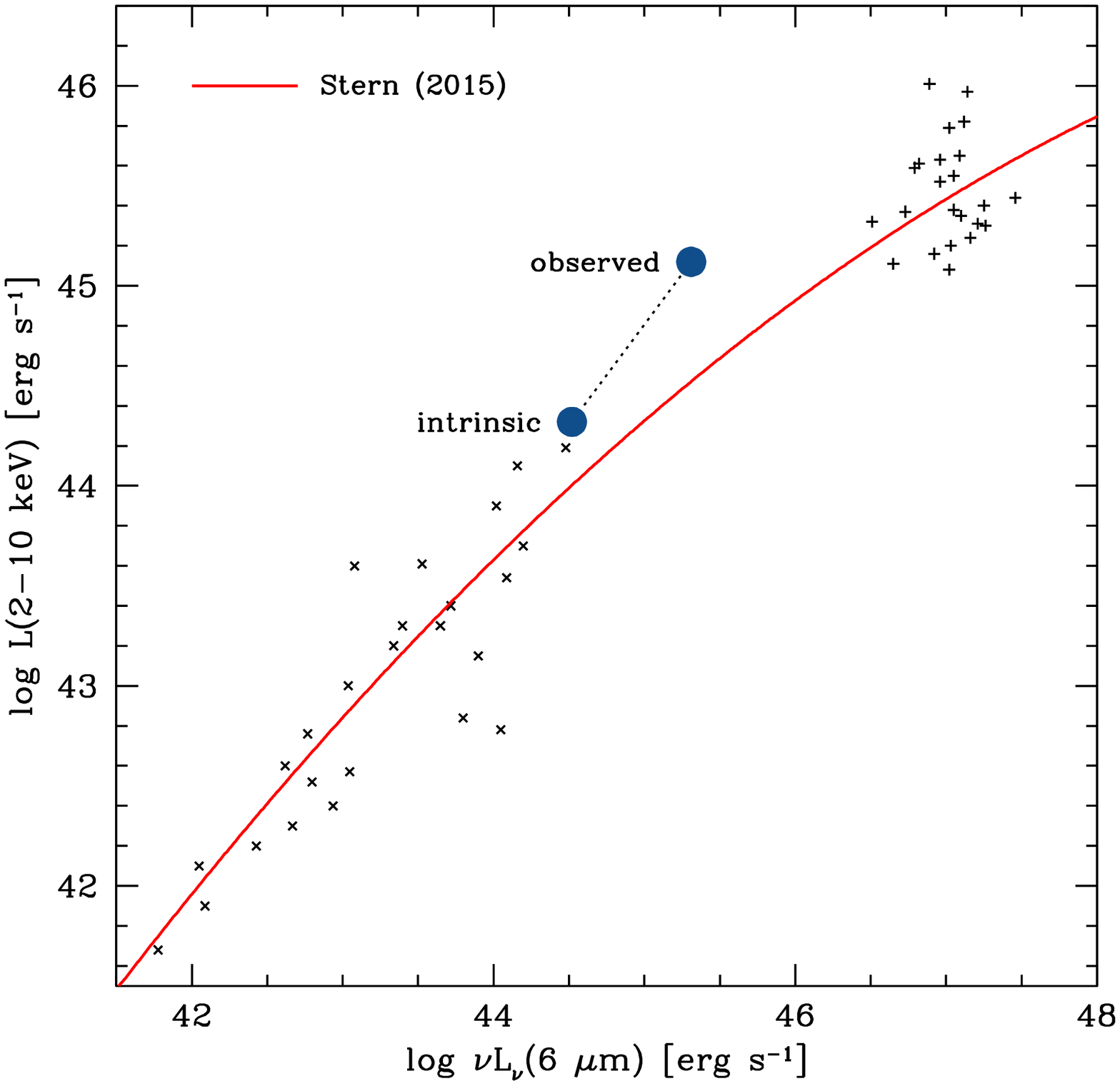}{3.2in}{0}{40}{40}{-120}{-50}
\caption{Rest-frame 2-10~keV X-ray luminosity against rest-frame 6$\mu$m luminosity for AGN over a wide range luminosity, with the relation from \citet{Stern:15} indicated as red solid curve. Black exes show local Seyfert galaxies from \citet{Horst:08} and \citet{Gandhi:09}, while black plus signs show luminous quasars from \citet{Just:07}. Large circles show observed and lensing-corrected values for \mg.  Intinsically, the radio galaxy lies close to observed relation, at a luminosity where the relation is roughly linear.  However, the observed value lies far above the relation, suggesting a new method to identify lensed AGN.}
\label{fig:LxLir}
\end{figure}

\section{Discussion}

With the redshift of \mg\ now known, we can do some rough calculations of the properties of the lensing galaxy.  Approximating the Einstein ring as a circle of diameter 1\farcs9, the radius of an Einstein ring, $\theta_{\rm E}$ (in radians) is related to the projected mass enclosed by the Einstein radius, $M_{\rm E}$ by
$$ \theta_{\rm E} = \sqrt{ {{4\, G\, M_{\rm E}}\over{c^2}}\, {{D_{\rm LS}}\over{D_{\rm S}\, D_{\rm L}}}},$$
where $D_{\rm S}$ ($D_{\rm L}$) is the angular diameter distance to the source (lens), and $D_{\rm LS}$ is the angular diameter distance between the source and lens. For a flat universe (i.e., $\Omega_{\rm M} + \Omega_\Lambda = 1$), $D_{\rm LS} = D_{\rm S} - D_{\rm L}\, (1 + z_l)/(1 + z_s)$.  For the \mg\ lens, we find $M_{\rm E} = 4.2 \times 10^{11}\, M_\odot$.  From {\it Hubble} near-infrared imaging, \citet{Kochanek:00b} find the lensing galaxy is well-fit by a de~Vaucouleurs profile with an effective radius $R_e = 0\farcs7$, which is much less than the Einstein radius.  Observed $I$-band roughly corresponds to rest-frame $B$-band for the lensing galaxy, so the observed $I$-band magnitude of the lensing galaxy, $I=21.04$ \citep{Kochanek:00b}, corresponds to a rest-frame $B$-band luminosity of $L_{\rm B} = 1.7 \times 10^{11}\, L_{\rm B, \odot}$.  The implied mass-to-light ratio of the lensing galaxy enclosed by the Einstein ring is then $M_{\rm E} / L_{\rm B} \approx 2.5\, M_\odot / L_{\rm B, \odot}$, which is low  for local early-type galaxies \citep[e.g.,][]{Gerhard:01}, but consistent with early-type galaxies at higher redshift \citep[e.g.,][]{vanderWel:05}.

Figure~\ref{fig:LxLir} shows rest-frame 2-10~keV X-ray luminosity against rest-frame 6$\mu$m luminosity for AGN over a wide range luminosity.  At low luminosities, the relation is roughly linear, while at high luminosities, the 2-10~keV X-ray emission essentially saturates as the corona more effectively cools and softens when the accretion disk thermal emission increases \citep[e.g.,][]{Brightman:13}.  Also shown on this figure are the observed and intrinsic luminosities of \mg. The mid-infrared luminosity is derived by extrapolating the observed {\it WISE} photometry.  {\it WISE} does not resolve the system, so this combines the flux from the lens and the source, though at these wavelengths, the early-type lensing galaxy is strongly overpowered by the AGN.  The X-ray luminosity is the sum of both point sources, to provide a consistent comparison with the {\it WISE} data.  With the source redshift now determined, \mg\ warrants a full lensing model, including sheer from galaxies close to the line of sight as well as an ellipsoidal model of the lensing galaxy. In lieu of a full model, which is beyond the scope of this paper, we instead adopt an approximate 2~mag magnification based on the optical and near-infrared magnifications in \citet{Kochanek:00}.  With this correction, \mg\ intrinsically resides close to the linear region of the X-ray-mid-infrared relation. Fig.~3 suggests a new way to identify candidate lensed AGN from low-resolution, wide-area X-ray surveys such as eROSITA, by targeting sources which have an anomalously high X-ray luminosity given their mid-infrared luminosity.

With the redshift of \mg, the first Einstein ring, now known, this opens up a range of follow-up observations and studies.  Given the moderate absorption found from the X-ray analysis, the expectation is that \mg\ will have broad Balmer emission lines, similar to reddened quasars \citep[e.g.,][]{Glikman:07b}. However, given the redshift of the source, this will require space-based observations.  Longer wavelength emission line studies are also made feasible with the source redshift known, such as targeted ALMA or APEX studies of the CO sled, as well as [\ion{C}{1}] and [\ion{C}{2}] emission.  Foremost, with the source redshift known, an updated lens model is called for. This would allow a range of studies, including improved measurements of the dark matter properties of the lensed galaxy.

% \citep[e.g.,][]{Nierenberg:20}.

\acknowledgements

We thank our colleagues, including Carlos De~Breuck, Anna Nierenberg, and Nick Seymour, for their input and discussions of this source.  In particular, we thank Leonidas Moustakas for providing access to the Master Lens Database.  The work of DS was carried out at the Jet Propulsion Laboratory, California Institute of Technology, under a contract with NASA. DJW acknowledges support from an STFC Ernest Rutherford Fellowship. Based on observations at the W. M. Keck Observatory, which is operated as a scientific partnership among the University of California, the California Institute of Technology, and the National Aeronautics and Space Administration. The Observatory was made possible by the generous financial support of the W. M. Keck Foundation.  The scientific results reported in this article are based in part on previously unpublished data obtained from the {\it Chandra} Data Archive.

\bibliographystyle{apj.bst}
% \bibliography{apj-jour,stern,mg1131_DJW}

\smallskip
{\it Facilities:} \facility{Chandra}, \facility{Keck (LRIS)}

\smallskip
\copyright 2020.  All rights reserved.

\clearpage
\end{document}